\documentclass[pra,twocolumn,showpacs,floatfix]{revtex4}
\usepackage{graphicx}
\usepackage{color}
\begin{document}

\title{Solitary-wave solutions in binary mixtures of Bose-Einstein
condensates under periodic boundary conditions}

\author{J. Smyrnakis$^1$, M. Magiropoulos$^1$, G. M. Kavoulakis$^1$,
and A. D. Jackson$^2$}
\affiliation{$^1$Technological Education Institute of Crete, P.O. 
Box 1939, GR-71004, Heraklion, Greece \\
$^2$The Niels Bohr International Academy, The Niels Bohr Institute,
 Blegdamsvej 17, DK-2100, Copenhagen \O, Denmark}
\date{\today}

\begin{abstract}

We derive solitary-wave solutions within the mean-field
approximation in quasi-one-dimensional binary mixtures of 
Bose-Einstein condensates under periodic boundary conditions, 
for the case of an effective repulsive interatomic interaction. 
The particular gray-bright solutions that give the global 
energy minima are determined. Their characteristics and 
the associated dispersion relation are derived. In the case 
of weak coupling, we diagonalize the Hamiltonian analytically 
to obtain the full excitation spectrum of ``quantum" 
solitary-wave solutions.  

\end{abstract}

 \maketitle

\section{Introduction}

The question of solitary-wave solutions in trapped 
atomic Bose-Einstein condensed gases has received 
considerable attention in recent years. Remarkably, 
solitary waves have been created and observed 
experimentally in both one-component \cite{burger, 
phillips, br1, br2} and two-component systems 
\cite{becker}. The easiest problem that one can 
consider is that of a quasi-one-dimensional Bose-Einstein 
condensate which extends to infinity with an effective 
repulsive interaction. In this case and within the 
mean-field approximation, the system is described 
by the nonlinear Schr\"odinger equation, which supports 
solitary-wave solutions, as was shown initially by Tsuzuki
\cite{Tsu}, and by Zakharov and Shabat \cite{ZS}. 

If one imposes periodic boundary conditions in a finite 
period of some fixed length, there are significant
differences. It is no longer possible to work in the 
thermodynamic limit, and it is necessary to impose the 
particle number normalization condition as well as the 
constraint of periodicity. This problem has been addressed 
within the mean-field approximation by Carr et 
al.\,\cite{carr}, for static solutions and by the 
authors of this study for the more general case of moving 
solutions \cite{SMKJ}. These were shown to be Jacobi 
elliptic functions. The excitation spectrum of the 
many-body problem of a single-component Bose gas 
interacting via a contact potential has also been 
studied by Lieb \cite{lieb}. This study assumed 
periodic boundary conditions and also considered the 
limit of an infinite period $L_0$ and an infinite 
number of particles $N$ with $N/L_0$ constant. The 
excitation spectrum was found to consist of two 
branches, one of which was later identified as 
corresponding to solitary-wave excitation \cite{id}. 

The problem of solitary-wave solutions in binary mixtures 
of Bose-Einstein condensates has also been studied 
theoretically by many people, see, e.g., \cite{trillo, 
christo, shalaby, santos, anglin, kevrekidis, berloff, 
konotop, liu}. In the case of strictly one-dimensional 
motion with the two condensates extending over an infinite 
line, the coupled system of two non-linear Schr\"odinger 
equations that describe the two order parameters (within 
the mean-field approximation) has been studied in the context 
of integrable systems by Manakov \cite{manakov}. 

In the present study we present mean-field solitary-wave 
solutions for a two-component Bose-Einstein condensate with 
repulsive interatomic interactions for a periodic system of 
finite length $L_0$.  To determine these solutions we impose 
constraints on the particle number for each component as well 
as the constraint of periodicity for each species. This 
formidable set of constraints is necessary to determine the 
solution and the dispersion relation, i.e., the energy versus 
the angular momentum, of this system.

As in the case of a single component, the solitary-wave solutions 
are again found to be Jacobi elliptic functions \cite{kavoulakis1}.  
In this case, however, the presence of the second component permits 
several qualitatively different solutions, i.e., ``gray-gray" and 
``gray-bright" solitons.  We will argue that the gray-bright 
solution has the lowest energy for particular choices of the 
phase winding numbers and for interactions which are weak in 
comparison with the kinetic energy of the atoms. The numerical 
solution of the equations which result from these constraints also 
suggests that the choice of the winding numbers does not change for 
moderate couplings. 

This problem is also intimately connected with the ``yrast" 
problem, i.e., with the evaluation of the state of lowest 
energy with some fixed angular momentum. The yrast state was 
determined in Refs.\,\cite{kavoulakis1, kavoulakis2} 
and coincides \cite{Ueda, equiv} with the gray-bright solution 
found below. The connection with the results of 
Refs.\,\cite{kavoulakis1, kavoulakis2}, obtained directly 
by minimization of the energy at fixed angular momentum,  
is clarified and gives further support to the belief that the yrast 
states are indeed the gray-bright solutions.  Although gray-gray 
solutions are also possible, the gray-bright solution has a 
lower energy.  The qualitative explanation of this result is simple.  
The density depression found in the gray component is (at least 
partially) filled by the bright component.  This leads to a more 
uniform total density and thus to a lower energy provided that 
the interatomic interaction is repulsive.

Finally, in the particular case of weak interatomic interactions, 
the complete energy spectrum of the system is determined by  
diagonalization the many-body Hamiltonian. This is done by 
identifying certain bilinears of the annihilation and creation 
operators with angular momentum operators. Under the assumption 
of weak coupling, the interaction terms in the Hamiltonian can 
be re-expressed in terms of Casimir operators and thus permit 
the analytic diagonaliztion of the Hamiltonian.  

In the following we first present our model in Sec.\,II and 
adopt an ansatz that allows us to solve the two coupled 
nonlinear Gross-Pitaevskii equations. In Sec.\,III we evaluate 
the energy and the angular momentum of these solutions. In Sec.\,IV 
we impose the constraints that are set by particle normalization 
and periodicity. In Sec.\,V we consider the nature of these solutions 
in the limit of weak interactions, and in Sec.\,VI we present  
numerical results for our solutions for stronger coupling. 
In Sec.\,VII we present the results from diagonalization of the 
many-body Hamiltonian.  Conclusions and an overview are given 
in Sec.\,VIII.

\section{Model and solitary-wave solutions}

Within the mean-field approximation, the order parameters 
of the two distinguishable species (labelled as $A$ and $B$) 
satisfy the coupled system of the following 
integrable non-linear equations (Manakov system),
\begin{eqnarray}
  i \frac {\partial \psi_A} {\partial t}
  &=& -\frac{1}{2}\psi_A''+(\gamma_{AA}|\psi_A|
  ^2+\gamma_{AB} |\psi_B|^2)\psi_A  
\label{eq:schr1} \\
  i \frac {\partial \psi_B} {\partial t} 
  &=& -\frac{1}{2}\psi_B''+(\gamma_{BA}|\psi_A|
 ^2+\gamma_{BB} |\psi_B|^2)\psi_B, 
\label{eq:schr2}
\end{eqnarray} 
where $\hbar=1$ and the masses of the two species are 
assumed to be equal (and are also set to unity). 
In addition $\gamma_{ij}$ is the matrix element for collisions
between species $i$ and $j$.

The solitary-wave solutions have the form of traveling waves 
with the particle density of each species moving with a constant 
velocity, $u$,
\begin{eqnarray}
   \psi_A= \sqrt{n_A(z)} e^{i\Phi_A(z)}e^{-i\mu_At} 
\label{eq:psi_A}\\
   \psi_B= \sqrt{n_B(z)} e^{i\Phi_B(z)}e^{-i\mu_Bt}, 
   \label{eq:psi_B}
\end{eqnarray}
where $z=x-ut$, $n_{A}$ and $n_B$ are the particle densities,
and $\mu_A$ and $\mu_B$ are the chemical potentials of the two 
species. Here $x$ is the spatial variable which is assumed to be periodic 
on the interval $0\le x \le L_0$. Following standard procedures, we 
separate the real and imaginary parts of these equations to find that 
\begin{eqnarray}
   \Phi_{A,B}' = u + \frac{C_{A,B}}{n_{A,B}},
\label{eq:PhiAB}
\end{eqnarray}
where $C_A$ and $C_B$ are constants of integration, and
also
\begin{eqnarray}
    \frac{1}{2} (\sqrt{n_A})'' = -\frac{1}{2}u^2 \sqrt{n_A}
    +\frac{1}{2} \frac{C_A^2}{n_A^{3/2}}
    \nonumber \\
+(\gamma_{AA} n_A + \gamma_{AB} n_B - \mu_A) \sqrt{n_A} 
\nonumber \\
\label{eq:RA}\\
    \frac{1}{2} (\sqrt{n_B})'' = -\frac{1}{2}u^2 \sqrt{n_B}
    +\frac{1}{2} \frac{C_B^2}{n_B^{3/2}}
    \nonumber \\
  +(\gamma_{BA} n_A + \gamma_{BB} n_B - \mu_B) \sqrt{n_B}.
\nonumber \\
\label{eq:RB}
\end{eqnarray}
Making the ansatz  
\begin{equation}
  n_B=\kappa n_A+\lambda,
\label{eq:ansatz}
\end{equation}
where $\kappa$ and $\lambda$ are parameters independent of
the space and time variables, we can integrate these equations. 
With this ansatz, the left side of Eqs.\,(\ref{eq:RA}) and 
(\ref{eq:RB}) become functions of $n_A$ and $n_B$ respectively 
that can be integrated to yield
\begin{eqnarray}
 \frac{1}{4}n_A'^2 = (\gamma_{AA}+\kappa\gamma_{AB})n_A^3 \hspace*{6.0em}
\nonumber \\
  - 2(\frac{1}{2}u^2+\mu_A-\lambda\gamma_{AB})n_A^2
   + E_An_A-C_A^2
\label{eq:nA}\\
    \frac{1}{4}n_B'^2=(\frac{\gamma_{BA}}{\kappa}
    +\gamma_{BB})n_B^3 \hspace*{6.0em}
\nonumber \\
    -2(\frac{1}{2}u^2+\mu_B+\frac{\lambda}{\kappa}\gamma_{BA})n_B^2
   +E_Bn_B-C_B^2,
\label{eq:nB}
\end{eqnarray}
where $E_A$ and $E_B$ are integration constants. Consistency 
of the ansatz translates into three equations that relate the 
integration constants and the chemical potentials of the two 
species together with the condition arising from the identification 
of the coefficients of $n_A^3$
\begin{equation}
\gamma_{AA}+\kappa\gamma_{AB}=\gamma_{AB}+\kappa\gamma_{BB}\equiv 
\gamma.
\label{condition}
\end{equation}
Our ansatz constrains the integration constants, thus
restricting the full solution space of our system of equations.  
In the generic case there are six integration constants 
in Eqs.\,(\ref{eq:nA}) and (\ref{eq:nB}), namely
$E_A, E_B, \mu_A, \mu_B, C_A$ and $C_B$ as well as the 
two constants $\kappa$ and $\lambda$ arising from the ansatz.  
There are also four consistency conditions which reduces the 
number of free constants to four for any given propagation velocity, 
$u$. The integration constant arising from the integration of 
Eqs.\,(\ref{eq:nA}) and (\ref{eq:nB}) is not included in this 
counting since it merely corresponds to a translation of the 
solution. However, the solution must also satisfy five constraints,  
namely two constraints of particle-number normalization, two 
phase-matching constraints, and one constraint which sets the 
period of the solution to $L_0$. In short, there are too many 
constraints.  

One possible way out of this dilemma would be to view the velocity 
of propagation, $u$, as a parameter to be set by the constraints.  
This, however, would lead to the unphysical result that the velocity 
of the waves cannot be changed without altering the properties of 
the atoms involved. If we ignore the ansatz, $u$ is expected to be 
a free parameter, since we have a total of six integration constants 
(including the chemical potentials) and six constraints (i.e., two 
particle number normalizations, two phase matchings and two density 
matchings.)  The restriction on $u$ must be viewed as an artifact 
of the ansatz.   

A more satisfactory way to deal with this problem is to fine tune 
the coupling constants.  If the masses of the two components 
are equal, it follows that $\gamma_{AB}$ and $\gamma_{BA}$ 
are trivially equal.  If, however, $\gamma_{AA} = \gamma_{AB} = 
\gamma_{BB}\equiv \gamma_0$ (i.e., if the scattering lengths between 
the same and the different species are all equal), then the condition 
of Eq.\,(\ref{condition}) becomes trivial, and we have five free 
constants and five constraints. This allows the velocity of propagation 
to be a free parameter. We thus proceed under the assumption that 
all the coupling constants are equal. Note that in this case 
Eq.\,(\ref{condition}) implies that $\gamma = \gamma_0 (1+\kappa)$.

It is convenient to factorize the right sides of Eqs.\,(\ref{eq:nA}) 
and (\ref{eq:nB}) to obtain the equations 
\begin{eqnarray}
   \frac{1}{2}n_A'^2=2\gamma (n_A-\rho_{A1})(n_A-\rho_{A2})
   (n_A-\rho_{A3})
\label{eq:nA2}\\
    \frac{1}{2}n_B'^2=\frac{2\gamma}{\kappa} (n_B-\rho_{B1})
    (n_B-\rho_{B2}) (n_B-\rho_{B3}),
    \label{eq:nB2}
\end{eqnarray}
where the roots in these equations are written in ascending order.  
Compatibility between Eqs.\,(\ref{eq:nA}) and (\ref{eq:nB}) and 
Eqs.\,(\ref{eq:nA2}), and (\ref{eq:nB2}) requires that 
\begin{eqnarray}
   C_A^2 &=& \gamma \rho_{A1} \rho_{A2} \rho_{A3} 
\label{eq:CA0} 
\\
C_B^2 &=& \frac{\gamma}{\kappa} \rho_{B1} \rho_{B2} \rho_{B3}. 
\label{eq:CB0} 
\end{eqnarray}
Since both densities have maximum and minimum values where their 
derivatives must vanish, all roots must be real.  Because of the 
positivity of  $n_A'^2$, the solution of  Eq.\,(\ref{eq:nA2}) is 
trapped between $\rho_{A1} =n_{A, {\rm min}}$ and $\rho_{A2}=
n_{A, \rm{max}}$, i.e., between the minimum and the maximum 
densities of species $A$. There are two possibilities for 
Eq.\,(\ref{eq:nB2}). For one of these, $\kappa>0$, $\rho_{B1}
=n_{B, {\rm min}}$, and $\rho_{B2} = n_{B, {\rm max}}$.  For 
the other, $\kappa<0$, $\rho_{B2}= n_{B, {\rm min}}$, and 
$\rho_{B3} = n_{B, {\rm max}}$. The first case corresponds 
to a gray-gray solution; the second corresponds to a 
gray-bright solution.  

For the gray-gray solution, we note that Eq.\,(\ref{eq:nB2}) 
reduces to Eq.\,(\ref{eq:nA2}) if  
\begin{eqnarray}
   n_{B, {\rm min}} &=& \kappa n_{A, {\rm min}}+\lambda 
\label{eq:GGmin} \\
   n_{B, {\rm max}} &=& \kappa n_{A, {\rm max}}+\lambda 
\label{eq:GGmax} \\
   \rho_{B3} &=& \kappa \rho_{A3}+\lambda. 
   \label{eq:GGrho3}
\end{eqnarray}
The solution of Eqs.\,(\ref{eq:nA2}), and (\ref{eq:nB2}) can 
then be expressed in terms of Jacobi elliptic functions as 
\begin{eqnarray}
   n_A=n_{A, {\rm min}}+(n_{A, {\rm max}}-n_{A, {\rm min}})
   {\rm sn}^2(\frac{2K(m)z}{L_0}|m)
\label{eq:GGnA} \\
   n_B=n_{B, {\rm min}}+(n_{B, {\rm max}}-n_{B, {\rm min}})
   {\rm sn}^2(\frac{2K(m)z}{L_0}|m), 
\label{eq:GGnB}
\end{eqnarray}
where
\begin{equation}
   m=\frac{n_{A, {\rm max}}-n_{A, {\rm min}}} {\rho_{A3}-
   n_{A, {\rm min}}}= \frac{n_{B, {\rm max}}-
  n_{B, {\rm min}}} {\rho_{B3}-n_{B ,{\rm min}}},
\label{eq:GGm}
\end{equation} 
and where the first elliptic integral $K(m)$ satisfies the 
periodicity constraint 
\begin{equation}
   K(m)=\frac{L_0}{\sqrt{8m}}\sqrt{2\gamma (n_{A, {\rm max}}
   -n_{A, {\rm min}})}.
\label{eq:periodicity}
\end{equation}
Equation (\ref{eq:GGrho3}) implies that $m$ is the same in 
Eqs.\,(\ref{eq:GGnA}) and (\ref{eq:GGnB}).  In this general 
form, the five independent constants in the solution space are 
$n_{A, {\rm max}}, n_{A, {\rm min}}, \kappa, \lambda$, and $m$. 

In the gray-bright case ($\kappa<0$), the reduction of 
Eq.\,(\ref{eq:nB2}) to Eq.\,(\ref{eq:nA2}) requires that 
\begin{eqnarray}
  n_{B, {\rm min}} &=& \kappa n_{A, {\rm max}} + \lambda 
\label{eq:GBmin} \\
  n_{B, {\rm max}} &=& \kappa n_{A, {\rm min}} + \lambda 
\label{eq:GBmax} \\
  \rho_{B1} &=& \kappa \rho_{A3}+\lambda. 
\label{eq:GBrho1}
\end{eqnarray}
The solution of Eqs.\,(\ref{eq:nA2}) and (\ref{eq:nB2}) is now
\begin{eqnarray}
   n_A=n_{A, {\rm min}}+(n_{A, {\rm max}}-n_{A, {\rm min}})
   {\rm sn}^2(\frac{2K(m)z}{L_0}|m) 
\label{eq:GBnA}\\
   n_B=n_{B, {\rm min}}+(n_{B, {\rm max}}-
   n_{B, {\rm min}}) {\rm cn}^2(\frac{2K(m)z}{L_0}|m), 
\label{eq:GBnB}
\end{eqnarray}
where
\begin{equation}
   m=\frac{n_{A, {\rm max}}-n_{A, {\rm min}}}{\rho_{A3}-
   n_{A, {\rm min}}}= \frac{n_{B, {\rm max}}-n_{B, {\rm min}}} 
   {n_{B, {\rm max}}-\rho_{B1}},
\label{eq:GBm}
\end{equation} 
and $K(m)$ satisfies the periodicity constraint given above.  
Again, the five independent constants in the solution space are 
$n_{A, {\rm max}}, n_{A, {\rm min}}, \kappa,\lambda$, and $m$. 

For both the gray-gray and the gray-bright cases, we note that 
Eqs.\,(\ref{eq:nA2}) and (\ref{eq:nB2}) have an interesting limiting 
form when $\kappa \to -1$ and $\rho_{A3} \to + \infty$ in such a 
way that $(1+\kappa )\rho_{A3}$ is finite.  The fact that $\rho_{A3} 
\to + \infty$ tells us that $m \to 0$.  Hence, the elliptic 
functions ${\rm sn}(2K(m)z/L_0|m)$, and ${\rm cn}(2K(m)z/L_0|m)$ 
become the regular trigonometric functions $\sin(\pi z/L_0)$ and 
$\cos(\pi z/L_0)$, respectively. The periodicity condition of 
Eq.\,(\ref{eq:periodicity}) assumes the form 
\begin{equation}
  \lim_{\kappa \to -1}\frac{m}{1+\kappa}=\frac{L_0^2}
{\pi^2}\gamma_0(n_{A, {\rm max}}-n_{A, {\rm min}}).
\label{degperiod}
\end{equation}  
The gray-gray solution then simplifies to 
\begin{eqnarray}
   n_A=n_{A, {\rm min}}+(n_{A, {\rm max}}-n_{A, {\rm min}})
   \sin^2(\frac{\pi z}{L_0})
\label{eq:degGGnA}\\
   n_B=n_{B, {\rm min}}+(n_{B, {\rm max}}-n_{B, {\rm min}})
   \sin^2(\frac{\pi z}{L_0}), 
\label{eq:degGGnB}
\end{eqnarray}
and the gray-bright solution becomes 
\begin{eqnarray}
   n_A=n_{A, {\rm min}}+(n_{A, {\rm max}}-n_{A, {\rm min}})
   \sin^2(\frac{\pi z}{L_0}) 
\label{eq:degGBnA}\\
   n_B=n_{B, {\rm min}}+(n_{B, {\rm max}}-n_{B, {\rm min}})
   \cos^2(\frac{\pi z}{L_0}). 
\label{eq:degGBnB}
\end{eqnarray}
Since the total density is more uniform in gray-bright case and since 
the interaction is repulsive, this solution will have a lower energy.  
The remainder of this paper will focus on the analysis of this solution 
only.

\section{Dispersion relation: energy versus angular momentum}

In this section we evaluate the energy and the angular momentum 
of the system in the gray-bright case.  We begin with the angular 
momentum, which has the form
\begin{eqnarray}
L &=& \frac{L_0}{2\pi}\int \psi_A^*(-i\frac{d}{dx})\psi_Adx+
\frac{L_0}{2\pi}\int \psi_B^*(-i\frac{d}{dx})\psi_Bdx 
\nonumber \\
&=& \frac{L_0}{2\pi}\int (n_A\Phi_A'+n_B\Phi_B')dx.
\label{eq:angular1}
\end{eqnarray} 
We can eliminate $\Phi_A'$ and $\Phi_B'$ using the equations for 
the conservation of particles, Eqs.\,(\ref{eq:PhiAB}), to obtain 
\begin{equation}
\frac{2\pi}{L_0}L=u(N_A+N_B)+(C_A+C_B)L_0,
\label{eq:angular2}
\end{equation}
where $N_A$ and $N_B$ are the particle numbers for the two 
species. 

The energy is given by 
\begin{equation}
E=\int \{ \frac{1}{2}|\psi_A^\prime |^2+\frac{1}{2}|\psi_B^\prime |^2 
+\frac{1}{2} \gamma_0 ( |\psi_A |^2+|\psi_B |^2 ) ^2  \} dx,
\label{eq:energy}
\end{equation} 
which can also be written as
\begin{eqnarray}
E=\int \frac 1 2 \{ \frac{(n_A^\prime )^2}{4n_A}+n_A(u+\frac{C_A}
{n_A})^2+\frac{(n_B^\prime )^2}{4n_B}
\nonumber \\
+n_B(u+\frac{C_B}{n_B})^2 + \gamma_0(n_A+n_B)^2 \} dx.
\label{eq:energy2}
\end{eqnarray}
Using Eqs.\,(\ref{eq:nA2})--(\ref{eq:CB0}) together with the 
particle number normalization, $E$ can be expressed in the form 
\begin{eqnarray}
E - E_0 = \frac{1}{2} u^2 (N_A+N_B)+u(C_A+C_B)L_0
\nonumber \\
-\frac{\gamma}{2}( 
(2-\frac{1}{m}) n_{A, {\rm min}}+ 
(1+\frac{1}{m})n_{A, {\rm max}}) N_A 
\nonumber \\
- \frac{\gamma}{2\kappa}( (2-\frac{1}{m})n_{B, {\rm max}}
+(1+\frac{1} {m})n_{B, {\rm min}}) N_B 
\nonumber \\
+\frac{\gamma}{2}(2n_{A, {\rm max}} n_{A, {\rm min}} + \frac {1}
{m} n_{A, {\rm max}}^2+(1-\frac{1}{m})n_{A, {\rm min}}^2)
L_0\nonumber \\
+\frac{\gamma}{2\kappa}(2n_{B, {\rm max}} n_{B, {\rm min}}+\frac{1}
{m}n_{B, {\rm min}}^2+(1-\frac{1}{m})n_{B, {\rm max}}^2)L_0,
\nonumber \\
\label{eq:energy3}
\end{eqnarray}
where 
\begin{equation}
E_0=\int_{-L_0/2}^{L_0/2} [ \frac{\gamma}{2}n_A^2+\frac{\gamma}
{2\kappa}n_B^2+\frac{\gamma_0}{2}(n_A+n_B)^2 ].
\label{eq:energy4}
\end{equation}
It is possible to derive a simple form for $E_0$ using the 
normalization constraints (see Appendix 2). In this way we get 
\begin{eqnarray}
E_0 = \gamma_0(1+\kappa )^2 \Big{[} n_{A, {\rm min}} (2N_A-
n_{A, {\rm min}} L_0
\nonumber \\
-\frac{2(m+1)}{3m}\Delta n_AL_0)+
\frac{2(m+1)}{3m} N_A \Delta n_A - \frac{1}{3m} \Delta 
n_A ^2L_0\Big{]} 
\nonumber  \\
+ \gamma_0\Big{[}\frac{N_AN_B}{L_0}-\frac{3}{2}\kappa\frac{N_A^2}
{L_0}-\kappa^2\frac{N_A^2}{L_0}+\frac{N_B^2}{L_0}+\frac{1}
{2\kappa}\frac{N_B^2}{L_0}\Big{]}.
\nonumber \\
\label{eq:energy5}
\end{eqnarray}

In the limiting case $\kappa \to -1$ with $(1+\kappa)= O(m)$, 
$E_0 \to \gamma_0 (N_A+N_B)^2/(2 L_0)$, which is indeed the 
interaction energy of a gas of constant density with $N_A+N_B$ 
particles. In this limit, the remaining energy is 
\begin{equation}
  E-E_0 \rightarrow \frac{1}{2}u^2N+u(C_A+C_B)L_0+
 \frac{1}{2}\frac{\pi^2} {L_0^2}N
\label{eq:degkinetic}.
\end{equation}
Here, we have used the limiting form of the periodicity condition, 
Eq.\,(\ref{degperiod}), to eliminate quotients of the form $(1+\kappa)
/m$. If the winding number $q_A=0$ or $q_B=0$, then $m=0$ and the 
velocity becomes $u=\pi/L_0$, as it should. Since we also have 
Eq.\,(\ref{eq:angular2}), it is possible to write 
\begin{equation}
  E-E_0\rightarrow \frac{2\pi^2}{L_0^2}L=\frac{u}{(L_0/2\pi)} L.
\label{eq:degdisp}
\end{equation}
This is consistent with the equation $\partial E/\partial L = 
\Omega$, where $\Omega$ is the angular velocity of the condensate.  

\section{Constraints}

The particle number normalization constraints tell us that
\begin{equation}
  \int_{-L_0/2}^{L_0/2}n_{A,B}dz = N_{A,B}.
\label{eq:normalization}
\end{equation} 
Using the integrals given in Appendix 2, these equations become
\begin{eqnarray}
   n_{A, {\rm min}} L_0+(n_{A, {\rm max}}-n_{A, {\rm min}})
   \frac{L_0}{m}(1-\frac{E(m)}{K(m)}) 
   \nonumber \\ = N_A 
\label{eq:normA} \\
    n_{B, {\rm min}} L_0+(n_{B, {\rm max}}-n_{B, {\rm min}})
    \frac{L_0}{m}(m-1+\frac{E(m)}{K(m)})  
    \nonumber \\ = N_B,
\label{eq:normB}
\end{eqnarray}
where $K(m)$ and $E(m)$ are the usual elliptic integrals.  

The phase matching constraints,  
\begin{eqnarray}
  \int_{-L_0/2}^{-L_0/2}\Phi_{A,B}' dz = 2\pi q_{A,B},
\label{pm}
\end{eqnarray}
imply that 
\begin{eqnarray}
uL_0+C_{A,B}\int_{-L_0/2}^{-L_0/2}\frac{1}{n_{A,B}}dz=2\pi q_{A,B}.
\label{eq:phasemach}
\end{eqnarray}
Here $q_A$ and $q_B$ are the winding numbers of the two species. 
Equations (\ref{eq:CA0}) and (\ref{eq:CB0}) allow us to determine
$C_A$ and $C_B$ as
\begin{eqnarray}
   C_A^2= \gamma 
   n_{A, {\rm min}} n_{A, {\rm max}}
   [n_{A, {\rm min}}+\frac{n_{A, {\rm max}}-n_{A, {\rm min}}}{m}]
\label{eq:CA}\\
   C_B^2= \frac{\gamma}{\kappa} 
   n_{B, {\rm min}} n_{B, {\rm max}}[n_{B, {\rm max}}
   -\frac{n_{B, {\rm max}}- n_{B, {\rm min}}}{m}].
\label{eq:CB}
\end{eqnarray}
Carrying out the integrations in Eq.\,(\ref{eq:phasemach}) 
and solving for the velocity $u$, we find that 
\begin{eqnarray}
  u=\frac{2\pi q_A}{L_0}
  \pm \sqrt{\frac{n_{A, {\rm max}}} {n_{A, {\rm min}}}}
  \frac 1 {K(m)} \Pi(1-\frac{n_{A, {\rm max}}}{n_{A, {\rm min}}}|m)
  \nonumber \\ \times
    \sqrt{\gamma[n_{A, {\rm min}}+ (n_{A, {\rm max}}-
n_{A, {\rm min}})/{m}]}
\nonumber \\
\label{eq:uA}
\end{eqnarray}
\begin{eqnarray}
  u=\frac{2\pi q_B}{L_0}
  \pm \sqrt{\frac{n_{B, {\rm min}}} {n_{B, {\rm max}}}}
  \frac 1 {K(m)} \Pi(1-\frac{n_{B, {\rm min}}}{n_{B, {\rm max}}}|m)
  \nonumber \\ \times
    \sqrt{(\gamma/\kappa)[n_{B, {\rm max}} - (n_{B, {\rm max}}-
n_{B, {\rm min}})/{m}]},
\label{eq:uB} 
\end{eqnarray}
where $\Pi(a|m)$ is the third elliptic integral. Note that the 
sign ambiguity that appears in Eqs.\,(\ref{eq:uA}) and (\ref{eq:uB}) 
comes from the ambiguity in the sign of the constants $C_A,C_B$.  
Equations (\ref{eq:nA}) and (\ref{eq:nB}) for the densities 
$n_A$ and $n_B$ involve only $C_{A,B}^2$, hence there is an 
ambiguity in the signs of $C_{A,B}$ in the solutions. The only 
place where these signs are important are in the phase matching 
constraints, where the winding numbers also appear. Therefore, a 
solution of the phase matching constraints involves not only a 
choice of the winding numbers $q_A$ and $q_B$, but also a choice 
for the signs of $C_{A,B}$. The final constraint is the periodicity 
constraint, Eq.\,(\ref{eq:periodicity}), encountered earlier. These 
five constraints are sufficient to determine the solution.

\section{Weak-coupling limit}

In the particular case $m\to 0$ the periodicity constraint 
of Eq.\,(\ref{eq:periodicity}) has, to lowest order, a 
particularly simple form (see Appendix 1): 
\begin{equation}
\frac{1+\kappa}{m}=\frac{\pi^2(1+m/2)}{L_0^2\gamma_0(
n_{A, {\rm max}}- n_{A, {\rm min}})}+{\cal O}(m^2).
\label{eq:degperiod}
\end{equation}
Since in general it is not necessarily true that 
$\Delta n_A\equiv n_{A, {\rm max}}-n_{A, {\rm min}} \ll 1/L_0$,
from the above equation follows that we should also take the 
limit $\kappa \to -1$, so that the value of $\lim_{(\kappa,m)
\to(-1,0)} (1+\kappa)/{m}$ is finite and determines $\Delta n_A$.

In this limit, the normalization constraints of Eqs.\,(\ref{eq:normA}) 
and (\ref{eq:normB}) can be written as  
\begin{eqnarray}
\frac{n_{A, {\rm max}}+n_{A, {\rm min}}}{2}+\frac{\Delta n_A}
{16}m+{\cal O}(m^2)=\frac{N_A} {L_0} 
\label{degnormA}\\
\frac{n_{B, {\rm max}}+n_{B, {\rm min}}}{2}-\frac{\Delta n_B}
{16}m+{\cal O}(m^2)=\frac{N_B}
{L_0}. 
\label{degnormB}
\end{eqnarray}  
Setting $\bar{n}_{A,B} \equiv (n_{A,B, {\rm max}}+
n_{A,B, {\rm min}})/{2}$, we see that $\bar{n}_{A,B}= 
N_{A,B}/{L_0}$ to lowest order in $m$.  

The phase constants $C_A$ and $C_B$ are given as 
\begin{eqnarray}
  C_A &=& \pm \frac{\pi}{L_0}
  \sqrt{n_{A, {\rm max}} n_{A, {\rm min}}}\sqrt{1+m\frac{\bar{n}_A}
{\Delta n_A}}
\nonumber \\
&=& \pm \frac{\pi}{L_0}\sqrt{n_{A, {\rm max}} n_{A, {\rm min}}}
(1+\frac{1} {2}m\frac{\bar{n}_A}{\Delta n_A}) 
\label{eq:degCA}\\
   C_B &=& \pm \frac{\pi}{L_0}\sqrt{n_{B, {\rm max}} n_{B, {\rm min}}}
   \sqrt{1-m\frac{\bar{n}_B} {\Delta n_B}}
\nonumber \\
&=& \pm \frac{\pi}{L_0}\sqrt{n_{B, {\rm max}} n_{B, {\rm min}}}
(1-\frac{1} {2}m\frac{\bar{n}_B}{\Delta n_B}). 
\label{eq:degCB}
\end{eqnarray}

Using the asymptotic expansion of $\Pi(a|m)$ and $K(m)$ (see 
Appendix 1), we see that the phase constraints of Eqs.\,(\ref{eq:uA}) 
and (\ref{eq:uB}) can be written as 
\begin{eqnarray}
u &=& \frac{\pi}{L_0}(2q_A\mp 1\mp \frac{m}{2}
\frac{\sqrt{n_{A ,{\rm max}}n_{A ,{\rm min}}}}
{\Delta n_{A}})+{\cal O}(m^2)
\label{eq:deguA} \\
u &=& \frac{\pi}{L_0}(2q_B\mp 1\pm \frac{m}{2}
\frac{\sqrt{n_{B ,{\rm max}}n_{B ,{\rm min}}}}
{\Delta n_{B}})+{\cal O}(m^2). 
\label{eq:deguB}
\end{eqnarray}
Here, the signs that appear in the above formulae are to be 
determined by the signs of the phase constants $C_A$ and $C_B$ 
(positive, upper sign; negative lower sign), which still need to 
be determined. We also note that these expansion formulae are 
valid only if $\Delta n_{A,B}/{\bar n}_{A,B} \gg m$ since the 
expansions of Eqs.\,(\ref{eq:degCA}) and (\ref{eq:degCB}) are 
not otherwise valid. 

It is also possible to expand the particle densities near $m=0$ 
by making use of the Lambert series of the Jacobi elliptic functions 
(see Appendix 1). This gives us  
\begin{eqnarray}
n_A(z)=n_{A ,{\rm min}}+\Delta n_A[(\frac{1}{2}-
\frac{1}{2}\cos(\frac{2\pi z}
{L_0}))
\nonumber \\
+\frac{m}{16}(1-\cos(\frac{4\pi z}{L_0}))]+O(m^2) 
\label{eq:degnA}\\
n_B(z)=n_{B ,{\rm min}}+\Delta n_B[(\frac{1}{2}+
\frac{1}{2}\cos(\frac{2\pi z}
{L_0}))
\nonumber \\
-\frac{m}{16}(1-\cos(\frac{4\pi z}{L_0}))]+O(m^2). 
\label{eq:degnB}
\end{eqnarray}

The angular momentum given by Eq.\,(\ref{eq:angular2}) has  
the expanded form 
\begin{eqnarray}
\frac{2\pi}{L_0}L=u(N_A+N_B)\pm \pi\sqrt{n_{A ,{\rm max}}
n_{A ,{\rm min}}}(1+\frac{1}
{2}m\frac{\bar{n}_A}{\Delta n_A})
\nonumber \\ \pm \pi \sqrt{n_{B ,{\rm max}}
n_{B ,{\rm min}}}
(1-\frac{1}{2}m\frac{\bar{n}_B}{\Delta n_B}),
\label{eq:degangular}
\end{eqnarray}
where it is again necessary to assume that $\Delta n_{A,B} \gg m$.  

Let us consider now the specific branch $(q_A,q_B)=(0,0)$. If the 
velocity $u>0$, then Eq.\,(\ref{eq:phasemach}) demands that $C_A<0$ 
and $C_B<0$; hence the lower signs in Eqs.\,(\ref{eq:deguA}) and 
(\ref{eq:deguB}) apply. The only way to realize this while still 
having a common velocity for the two species is to have $m=0$ 
exactly.  This gives a propagation velocity of $u={\pi(2q_A+1)}/
{L_0}={\pi}/{L_0}$. This means that the densities have the form 
\begin{eqnarray}
n_A(z)&=&n_{A ,{\rm min}}+\Delta n_A(\frac{1}{2}-
\frac{1}{2}\cos(\frac{2\pi z}
{L_0})) 
\label{eq:degnA2}
\\
n_B(z)&=&n_{B ,{\rm min}}+\Delta n_B(\frac{1}{2}+
\frac{1}{2}\cos(\frac{2\pi z}
{L_0})). 
\label{eq:degnB2}
\end{eqnarray}
In this case the angular momentum is  
\begin{eqnarray}
\frac{2\pi}{L_0}L=\frac{\pi}{L_0}(N_A+N_B)- 
\pi \sqrt{n_{A ,{\rm max}}n_{A ,{\rm min}}}- 
\nonumber \\
\pi \sqrt{n_{B ,{\rm max}}n_{B ,{\rm min}}}.
\label{eq:degangular2}
\end{eqnarray}
Recalling that $n_{A ,{\rm max}}=-n_{B ,{\rm min}}+N/L_0$ and 
that $n_{A ,{\rm min}}=- n_{B ,{\rm max}}+N/L_0$, it is easy to 
show that the maximum value of the angular momentum in this branch 
is 
\begin{equation}
  \frac{L_{\rm max}}{N}\equiv \ell_{\rm max} =
  \frac{1}{2}-\frac{1}{2}\sqrt{x_A-x_B}.
\label{eq:maxl}
\end{equation}
Making use of the fact that $(n_{B, {\rm min}} + 
n_{B, {\rm max}})/2 = N_B/L_0 + O(m \Delta n_B)$ we note 
that this maximum is attained when $n_{B, {\rm min}}
n_{B, {\rm max}}$ vanishes and this is only possible when
$n_{B, {\rm min}} = 0$. Here, we have written the total 
particle number as $N=N_A+N_B$, the angular momentum per 
particle as $\ell=L/N$, and the particle fractions as 
$x_{A,B}=N_{A,B}/N$. As mentioned earlier, the yrast state 
of a two-component Bose-Einstein condensate confined in a 
ring trap was evaluated in Ref.\,\cite{kavoulakis2}. Based
on rather general arguments \cite{Ueda,equiv}, the present 
calculation is expected to be equivalent to that of the yrast 
state. Indeed, the branch found above corresponds to a portion 
of the first linear branch in the dispersion relation determined 
in Ref.\,\cite{kavoulakis2} (i.e., for $\ell < x_B$). The
lowest-energy state obtained in Ref.\,\cite{kavoulakis2} was 
found to be
\begin{eqnarray}
  \Psi_A & = & \sqrt{\frac{N_A}{L_0}}(c_0+c_1e^{2\pi i z/L_0}) 
\nonumber \\
   \Psi_B & = & \sqrt{\frac{N_B}{L_0}}(d_0+d_1e^{2\pi i z/L_0}),
\label{eq:oldwave}
\end{eqnarray}
where 
\begin{eqnarray}
  |c_0|^2=\frac{(x_A-\ell)(1-\ell)}{x_A(1-2\ell)}, && |c_1|^2=
\frac{\ell(x_B-\ell)} {x_A(1-2\ell)} 
\nonumber \\
  |d_0|^2=\frac{(x_B-\ell)(1-\ell)}{x_B(1-2\ell)}, && |d_1|^2=
\frac{\ell(x_A-\ell)} {x_B(1-2\ell)}
\label{eq:amplitudes}
\end{eqnarray}
for this branch. This gives rise to the following densities for
the two species 
\begin{eqnarray}
 n_A&=&\frac{N_A}{L_0}+\frac{N_A}{L_0}\left(c_0c_1^*e^{-2\pi i 
  z/L_0}+c_0^*c_1 e^{2\pi i z/L_0}\right)
\nonumber \\
  n_B&=&\frac{N_B}{L_0}+\frac{N_B}{L_0}\left(d_0d_1^*e^{-2\pi i 
  z/L_0}+d_0^*d_1e^{2\pi i z/L_0}\right).
\end{eqnarray}
These become identical to the densities given by Eqs.\,(\ref{eq:degnA2}) 
and (\ref{eq:degnB2}) if we set 
\begin{eqnarray}
c_0c_1^*=-\frac{n_{A ,{\rm max}}-n_{A ,{\rm min}}}{4N_A/L_0}, 
 d_0d_1^*=\frac{n_{B ,{\rm max}}-
n_{B ,{\rm min}}}{4N_B/L_0}. 
\label{eq:ident}
\end{eqnarray}
This is compatible with the amplitudes of Eqs.\,(\ref{eq:amplitudes}).  

We consider now the branch $(q_A,q_B)=(0,1)$. It will be 
assumed that the particle wave functions switch continuously to 
this branch as the angular momentum of the system increases. When 
applied to the phase matching constraints of Eqs.\,(\ref{eq:phasemach}), 
this continuity demands that $C_A<0$ and $C_B>0$.  To ${\cal O}(m)$ 
the phase matching condition can be satisfied in two ways. One is 
by having $m=0$, in which case the densities are exactly as in the 
$(q_A,q_B)=(0,0)$ case.  However, the constant $C_B$ now changes 
sign and becomes positive. This means that the angular momentum 
is given as 
\begin{eqnarray}
\frac{2\pi}{L_0}L=\frac{\pi}{L_0}(N_A+N_B)- \pi
\sqrt{n_{A ,{\rm max}}n_{A ,{\rm min}}}
\nonumber \\
+ \pi \sqrt{n_{B ,{\rm max}}n_{B ,{\rm min}}}.
\label{eq:degangular3}
\end{eqnarray}
This satisfies the inequality 
\begin{equation}
\frac{1}{2}-\frac{1}{2}\sqrt{x_A-x_B}\le \ell \le x_B.
\label{eq:boundl}
\end{equation}
This branch corresponds to the remainder of the first linear 
branch in the dispersion relation of Ref.\,\cite{kavoulakis2}.  

The second way is by having 
\begin{equation}
  \frac{\sqrt{n_{A ,{\rm max}}n_{A ,{\rm min}}}}{\Delta 
    n_A}=\frac{\sqrt{n_{B ,{\rm max}}n_{B ,{\rm min}}}}{\Delta n_B}
\label{eq:degmatch}
\end{equation}
to ${\cal O}(1)$. This relation together with the ansatz relations 
Eqs.\,(\ref{eq:GBmin}) and (\ref{eq:GBmax}) and the normalization 
conditions Eqs.\,(\ref{degnormA}) and (\ref{degnormB}) to lowest 
order tell us that $\kappa=-{N_B}/{N_A}$ and $\lambda={2N_B}/
{L_0}$. This is to be understood as a weak-coupling branch: 
The periodicity condition of Eq.\,(\ref{eq:degperiod}) demands 
that $\Delta n_A={\cal O}(m)$, and this violates the condition 
$\Delta n_{A,B}/ {\bar n}_{A,B} \gg m$, unless $\gamma_0 \ll {1}/
{L_0}$ is small. The angular momentum has the form 
\begin{equation}
  L=\frac{N_A+N_B}{2}-\frac{1}{2}\frac{\sqrt{n_{A ,{\rm max}}
 n_{A ,{\rm min}}}}{N_A/L_0} (N_A-N_B).
\label{eq:degangular4}
\end{equation}
Dividing by the total number of particles $N$, the above equation 
gives 
\begin{equation}
  \frac{\sqrt{n_{A ,{\rm max}}n_{A ,{\rm min}}}}{N_A/L_0}=
  \frac{1-2\ell}{x_A-x_B},
\label{geommean}
\end{equation}
and the propagation velocity becomes
\begin{equation}
  u=\frac{\pi}{L_0}+\frac{\gamma_0N}{2\pi}(1-2\ell).
\label{eq:degu}
\end{equation}
The minimum and the maximum values of the above angular momentum 
are given by the inequality 
\begin{equation}
  x_B \le \ell \le \frac{1}{2}.
\label{eq:boundl2}
\end{equation}
This branch can be identified as the first half of the curved part 
of the dispersion relation derived in Ref.\,\cite{kavoulakis2}. 
Note that at $\ell=1/2$ we get $u=\pi/L_0$, which is the velocity 
when $m$ is exactly zero.  

The next branch which appears as the angular momentum increases 
is given by $(q_A,q_B)=(1,0)$. Here we have $C_A>0$ and $C_B<0$. 
To order ${\cal O}(m)$ the phase matching conditions of 
Eqs.\,(\ref{eq:deguA}) and (\ref{eq:deguB}) may be satisfied in 
two ways.

One is by again demanding the validity of Eq.\,(\ref{eq:degmatch}) 
to order ${\cal O}(1)$.   As before, this leads to $\kappa=-{N_B}/
{N_A}$ and to the angular momentum 
\begin{equation}
  L=\frac{N_A+N_B}{2}+\frac{1}{2}\frac{\sqrt{n_{A ,{\rm max}}
 n_{A ,{\rm min}}}}{N_A/L_0} (N_A-N_B).
\label{eq:degangular5}
\end{equation}
Hence, 
\begin{equation}
  \frac{\sqrt{n_{A ,{\rm max}}n_{A ,{\rm min}}}}{N_A/L_0}
 =\frac{2\ell-1}{x_A-x_B},
\label{geommeann}
\end{equation}
and the propagation velocity is given again by Eq.\,(\ref{eq:degu}).   
However, the velocity is now lower than that for $m=0$, i.e., 
$u=\pi/L_0$. The minimum and the maximum values of the angular 
momentum are given by the inequality 
\begin{equation}
   \frac{1}{2}\le \ell \le x_A.
\label{eq:boundl3}
\end{equation}
This gives the second half of the curved part of the dispersion 
relation evaluated in Ref.\,\cite{kavoulakis2}. 

The other possibility is to set $m=0$. In this case the angular 
momentum becomes 
\begin{eqnarray}
  \frac{2\pi}{L_0}L=\frac{\pi}{L_0}(N_A+N_B)+ 
\pi\sqrt{n_{A ,{\rm max}}n_{A ,{\rm min}}}
\nonumber \\
- \pi \sqrt{n_{B ,{\rm max}}n_{B ,{\rm min}}},
\label{eq:degangular6}
\end{eqnarray}
and it satisfies the inequality 
\begin{equation}
   x_A\le \ell \le \frac{1}{2}+\frac{1}{2}\sqrt{x_A-x_B}.
\label{eq:boundl4}
\end{equation}
This reproduces a portion of the second linear branch in the 
dispersion relation evaluated in Ref.\,\cite{kavoulakis2}.

Finally the branch $(q_A,q_B)=(1,1)$ appears. Here, we necessarily 
have $m=0$, and the angular momentum has the form (with $C_A>0$ 
and $C_B>0$)
\begin{eqnarray}
  \frac{2\pi}{L_0}L=\frac{\pi}{L_0}(N_A+N_B)+ 
  \pi\sqrt{n_{A ,{\rm max}}n_{A ,{\rm min}}}
\nonumber \\
+ \pi \sqrt{n_{B ,{\rm max}}n_{B ,{\rm min}}}.
\label{eq:degangular7}
\end{eqnarray}
In this case the minimum value of the angular momentum is 
\begin{equation}
\ell_{\rm min}=\frac{1}{2}+\frac{1}{2}\sqrt{x_A-x_B}, 
\label{eq:minl}
\end{equation}
and the maximum value is $\ell=1$. This describes the remainder of 
the linear part of the dispersion relation of Ref.\,\cite{kavoulakis2}, 
in the interval $0<\ell<1$. Beyond $\ell=1$ the picture repeats itself 
because of Bloch's theorem \cite{Bl}, which tells us that an 
increase of $\ell$ by an integer can be attributed to excitation of 
the center of mass motion.  

\section{Numerical solution of the constraints}

It is possible to find numerical solutions to the constraint 
equations in order to determine $\kappa$ and $m$. The constraints 
that are crucial in determining these parameters are the phase 
constraints of Eq.\,(\ref{eq:phasemach}). If we restrict ourselves 
initially to the branch $(q_A,q_B)=(0,1)$, we first express $u$ in 
terms of the angular momentum per particle $\ell$ and obtain 
$\kappa$ and $m$ as functions of the angular momentum $\ell$. This 
solution is then readily extended to the branches $(q_A,q_B)=(0,0),
(0,1)$ and $(1,1)$, enabling us to plot various observables of the 
solitary waves as functions of $\ell$.

\begin{figure}
\includegraphics[width=9.5cm,height=10.5cm]{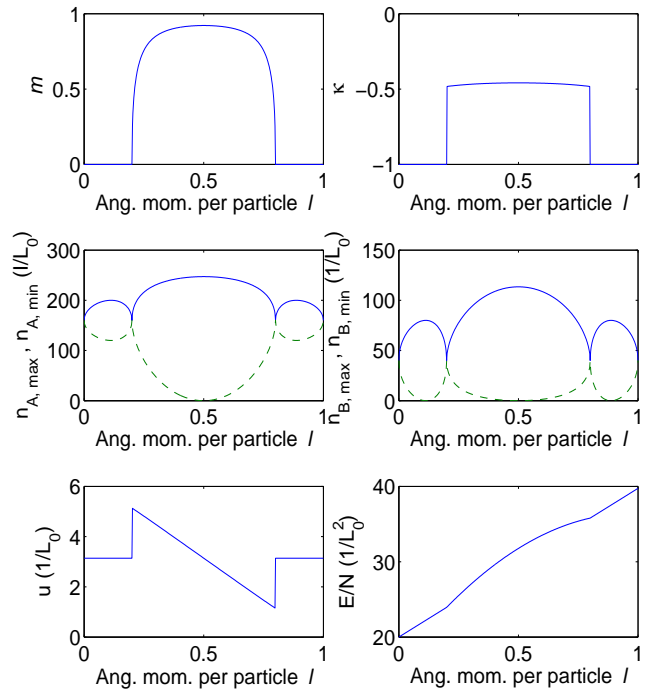}
\caption{The parameters $m$, $\kappa$, $n_{A,B, {\rm max}}$
(solid line), $n_{A,B, {\rm min}}$ (dashed line), the 
propagation velocity and the dispersion relation as 
functions of the angular momentum per particle $l$, 
for $N_A=160$, $N_B=40$, and $\gamma_0 L_0 = 0.2$. For 
$\ell=1/2$ there is a node in the density of each species 
(at different points). Also, the velocity of propagation 
$u$ has a discontinuity at $\ell = x_A$ and $\ell = x_B$.}
\end{figure}

One interesting aspect of the periodic gray-bright solution 
with period equal to $L_0$, which is expected to be the yrast 
state (i.e., the state of minimum energy for some fixed value 
of the angular momentum), is its size relative to $L_0$. A 
reasonable measure of this size is the ratio of the complex 
to the real period of the doubly periodic Jacobi solution, 
since the complex period controls the exponential decay of 
the solution. In Fig.\,2 we plot the ratio of the two periods 
versus the angular momentum. This ratio becomes infinite 
when $\ell \to N_B/N$, and it has its minimum value when 
$\ell=1/2$. This suggests that the solution is most localized when 
$\ell=1/2$.  However, even in this case the period ratio is not 
close to zero, suggesting that the yrast state is not very 
localized but rather has a size comparable to $L_0$ even 
for strong interatomic interactions. As seen from Fig.\,2 
when $\gamma_0 L_0 = 0.2$, i.e., when the ratio between the 
interaction energy of the homogeneous system and the kinetic 
energy, $N \gamma_0 L_0/(4 \pi^2)$, is equal to 1, the
minimum value of the ratio of the periods is $\approx 0.6$.
We have also found numerically that the corresponding minimal 
ratio for $\gamma_0 L_0 = 2$, is $\approx 0.37$, suggesting 
that the size of the waves is comparable to $L_0$ even for 
strong coupling.
 
\begin{figure}
\includegraphics[width=9.5cm,height=8.5cm]{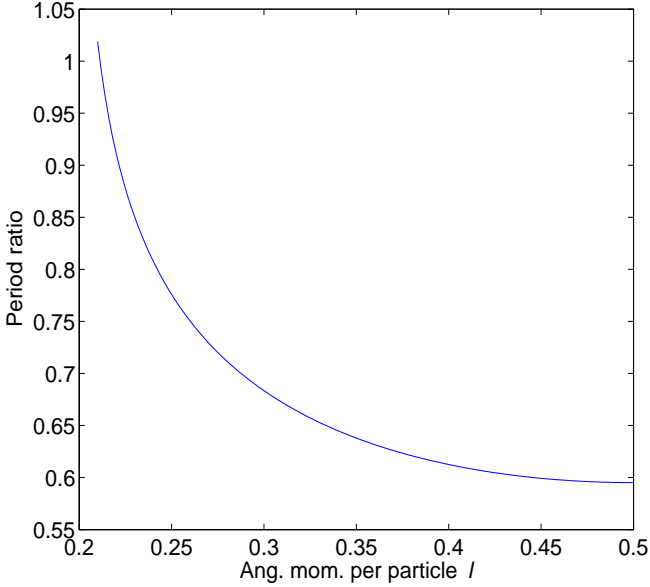}
\caption{The ratio of the imaginary to the real period of 
the gray-bright solution for $N_A=160$, $N_B=40$, and $\gamma_0 
L_0 = 0.2$ as a function of $l$. The period ratio diverges 
to infinity as $\ell \to N_B/N=0.2$. When $\ell=1/2$ we have the 
minimal ratio, suggesting that we have the most localized solution. 
However, the ratio is $\approx 0.6$ and thus the size of the 
localized wave is still approximately two thirds of the size 
of the period length $L_0$.}
\label{Fig:cohlength}
\end{figure}

\section{Diagonalization of the Hamiltonian for weak interactions}
 
For sufficiently weak interactions, $\gamma_0L_0<<1$, it is 
reasonable to truncate our Hamiltonian to the lowest two angular 
momentum modes only. Doing this, the second-quantized Hamiltonian 
becomes 
\begin{eqnarray}
\widehat{H} = \frac{1}{2}\left( \frac{2\pi}{L_0} \right)^2
(a_1^{\dagger }a_1+b_1^{\dagger }b_1) \hspace*{8.0em}
\nonumber \\
+ \frac{1}{2} \frac{\gamma_0}{L_0} (a_0^{\dagger 2}a_0^2+
b_0^{\dagger 2}b_0^2+
a_1^{\dagger 2}a_1^2+b_1^{\dagger 2}b_1^2 
\nonumber \\
+4a_0^{\dagger }a_1^{\dagger }a_0a_1+
4b_0^{\dagger }b_1^{\dagger} b_0b_1 \hspace*{5.5em}\nonumber \\
+ 2a_0^{\dagger }b_0^{\dagger }a_0b_0+2a_0^{\dagger } 
b_1^{\dagger }a_0b_1+
2a_1^{\dagger }b_0^{\dagger }a_1b_0
\nonumber \\
+2a_1^{\dagger }b_1^{\dagger }
a_1b_1 + 2a_1^{\dagger }b_0^{\dagger }a_0b_1+2a_0^{\dagger }
b_1^{\dagger }a_1b_0). 
\label{eq:quanthamilt}
\end{eqnarray}
We can diagonalize this Hamiltonian by considering the algebra of 
the bilinears of annihilation-creation operators appearing in it. 
We define the operators 
\begin{eqnarray}
\widehat{n}_{A0}&=&a_0^{\dagger }a_0, \,\,
\widehat{n}_{B0}=b_0^{\dagger }b_0, \,\, 
\widehat{n}_{A1}=a_1^{\dagger }a_1,  \,\, 
\widehat{n}_{B1}=b_1^{\dagger }b_1,  \nonumber \\
\widehat{J}_A&=&a_1^{\dagger }a_0, \,\, 
\widehat{J}_A^{\dagger }=a_0^{\dagger }a_1, \,\, 
\widehat{J}_B=b_1^{\dagger }b_0, \,\, 
\widehat{J}_B^{\dagger }=b_0^{\dagger }b_1. 
\label{eq:bilinears}
\end{eqnarray}  
The first four operators are the usual number operators. 
The last four can be used to generate two copies of the 
${\rm SU}(2)$ algebra. Since $[\widehat{n}_{A1}- 
\widehat{n}_{A0} , \widehat{J}_A] = 2 \widehat{J}_A$, 
it is natural to define 
\begin{eqnarray}
\widehat{J}_{A3}=\frac{1}{2}(\widehat{n}_{A1}-\widehat{n}_{A0}), & 
\widehat{J}_{B3}=\frac{1}{2}(\widehat{n}_{B1}-\widehat{n}_{B0}).
\label{eq:jay3}
\end{eqnarray}
These bilinears can be divided into three sets. One set 
consists of the operators $\widehat{n}_{A0}+\widehat{n}_{A1}$ 
and $\widehat{n}_{B0}+\widehat{n}_{B1}$. These operators 
commute with all other bilinears and are hence central elements.  
Their eigenvalues are determined by the particle numbers $N_A$ 
and $N_B$. The second set consists of the operators 
$\widehat{J}_{A3}$, $\widehat{J}_A$, and $\widehat{J}_A^{\dagger }$.
The third set consists of the operators $\widehat{J}_{B3}$, 
$\widehat{J}_B$, and $\widehat{J}_B^{\dagger }$.  The operators in 
the second set commute with the operators in the third set and the 
operators in each set satisfy the commutation relations 
\begin{eqnarray}
 [\widehat{J}_{A3,B3},\widehat{J}_{A,B}]=\widehat{J}_{A,B}, \,\, 
 [\widehat{J}_{A3,B3}, \widehat{J}_{A,B}^{\dagger }] = - 
 \widehat{J}_{A,B}^{\dagger}, 
\end{eqnarray}
and
\begin{eqnarray}
  [\widehat{J}_{A,B}, \widehat{J}_{A,B}^{\dagger}] = 2 
\widehat{J}_{A3,B3}.
\label{crcc}
\end{eqnarray}
These are the ${\rm SU}(2)$ commutation relations, which means 
that the algebra of the bilinears splits into a direct sum of two 
copies of the ${\rm U}(1)$ algebra and two copies of the ${\rm SU}(2)$ 
algebra. Note the angular momentum operator can be expressed in 
terms of these operators as
\begin{equation}
  \widehat{L}=\widehat{n}_{A1}+\widehat{n}_{B1}=\frac{1}{2}
 (N_A+N_B)+(\widehat{J}_{A3}+\widehat{J}_{B3}).
\label{eq:angoperator}
\end{equation} 
It is now possible to split the Hamiltonian into a central part 
and an ${\rm SU}(2)$ part, $\widehat{H} = H_C + \widehat{H}_0$, 
where 
\begin{eqnarray}
H_C=\frac{\pi^2}{L_0^2}(N_A+N_B)-\frac{1}{2}\frac{\gamma_0}{L_0}
(N_A+N_B)
\nonumber \\
+\frac{3}{4}\frac{\gamma_0}{L_0}(N_A^2+N_B^2)
+\frac{\gamma_0}{L_0}N_AN_B,
\label{eq:hamcentral}
\end{eqnarray}
and 
\begin{equation}
\widehat{H}_0=\frac{1}{2}\left( \frac{2\pi}{L_0} 
\right)^2(\widehat{J}_{A3}+\widehat{J}_{B3})+\frac{\gamma_0}{L_0}
(\widehat{J}_A\widehat{J}_B^{\dagger } + \widehat{J}_A^{\dagger} 
\widehat{J}_B-\widehat{J}_{A3}^2-\widehat{J}_{B3}^2).
\label{eq:hamzero}
\end{equation}
The linear part of $\widehat{H}_0$ is can readily be expressed in 
terms of the angular momentum.  Since the Hamiltonian is rotationally 
symmetric, the quadratic part of $\widehat{H}_0$ must commute with 
the angular momentum. This places significant constraints on the form 
of the quadratic part of the Hamiltonian. Indeed, if we set 
$\widehat{J}_{A,B} = \widehat{J}_{A1,B1} + i \widehat{J}_{A2,B2}$ 
and $\vec{\widehat{J}}_{A,B}=(\widehat{J}_{A1,B1},\widehat{J}_{A2,B2},
\widehat{J}_{A3,B3})$, it is possible to rewrite $\widehat{H}_0$ in 
the form 
\begin{eqnarray}
\widehat{H}_0=\frac{1}{2}\left( \frac{2\pi}{L_0} \right)^2
(\widehat{L}-N/2)-\frac{\gamma_0}{L_0}(\widehat{L}-N/2)^2
\nonumber \\
+\frac{\gamma_0}{L_0}[(\vec{\widehat{J}}_{A} + 
\vec{\widehat{J}}_{B})^2 
- \vec{\widehat{J}}_A^2-\vec{\widehat{J}}_B^2].
\label{hamzero2}
\end{eqnarray}
Since the angular momentum $\widehat{L}$ depends only on 
$\widehat{J}_{A3}+\widehat{J}_{B3}$, it commutes with 
$\vec{\widehat{J}}_{A,B}^2$ and with $(\vec{\widehat{J}}_{A} 
+\vec{\widehat{J}}_{B})^2$. Also, since $\widehat{J}_{A3,B3}$ 
are given in terms of the number operators in Eq.\,(\ref{eq:jay3}), 
it is clear that their eigenvalues range from $-N_{A,B}/2$ to 
$N_{A,B}/2$. This means that we are in the spin $j_{A,B}=N_{A,B}/2$ 
representation of the ${\rm SU}(2)$ algebra. This means that 
$\vec{\widehat{J}}_{A,B}^2$ and $(\vec{\widehat{J}}_{A} + 
\vec{\widehat{J}}_{B})^2$ are given by 
\begin{eqnarray}
\vec{J}_{A,B}^2 &=& \frac{N_{A,B}}{2}\left( \frac{N_{A,B}}{2}+1 
\right)\widehat{I} 
\\
(\vec{J}_{A}+\vec{J}_{B})^2 &=& j_{AB}(j_{AB}+1)\widehat{I},
\label{eq:casimirAB}
\end{eqnarray}
where $\widehat{I}$ stands for the identity operator and $j_{AB}$ 
ranges from $|j_A-j_B|=(N_A-N_B)/2$ to $j_A+j_B=N/2$ as a consequence 
of the usual rules for the addition of angular momentum.  The 
eigenvalues of the angular momentum operator $\widehat{L}$ are 
$L=N/2+m_{AB}$, where $m_{AB}$ ranges from $-j_{AB}$ to $j_{AB}$ due 
to Eq.\,(\ref{eq:angoperator}). Therefore, $\widehat{H}_0$ can be 
written in the form 
\begin{eqnarray}
\widehat{H}_0=\frac{1}{2}\left( \frac{2\pi}{L_0} \right)^2
(\widehat{L}-N/2)-\frac{\gamma_0}{L_0}(\widehat{L}-N/2)^2 
\hspace*{7.0em}
\nonumber \\
+\frac{\gamma_0}{L_0}\left[j_{AB}(j_{AB}+1)-\frac{N_{A}}{2}
\left( \frac{N_{A}}{2}+1 \right)-
\frac{N_{B}}{2}\left( \frac{N_{B}}{2}+1 \right)\right]\widehat{I},
\nonumber \\
\label{hamzero3}
\end{eqnarray}
and its eigenvalues are 
\begin{eqnarray}
E_0=\frac{1}{2}\left( \frac{2\pi}{L_0} \right)^2(L-N/2)-
\frac{\gamma_0}{L_0}(L-N/2)^2
\nonumber \\
+\frac{\gamma_0}{L_0}[j_{AB}(j_{AB}+1)-\frac{N_{A}}{2}
\left( \frac{N_{A}}{2}+1 \right)-
\frac{N_{B}}{2}\left( \frac{N_{B}}{2}+1 \right)].
\nonumber \\
\label{ezero}
\end{eqnarray}
Adding to this the contribution from $H_C$ we get that the energy 
eigenvalues are
\begin{eqnarray}
E = -\frac{\gamma_0}{L_0}L^2+\frac{1}{2}\left( \frac{2\pi}{L_0} 
\right)^2L+\frac{\gamma_0}{L_0}NL+
\frac{1}{2}\frac{\gamma_0}{L_0}N(N-1)
\nonumber \\
-\frac{1}{2}\frac{\gamma_0} {L_0}N_AN_B 
+\frac{\gamma_0}{L_0}[j_{AB}(j_{AB}+1)-\frac{N_{A}}{2}\left( 
\frac{N_{A}}{2}+1 \right)
\nonumber \\
-\frac{N_{B}}{2}\left( \frac{N_{B}}{2}+1 \right)].
\nonumber \\
\label{eq:enereigen}
\end{eqnarray}
It is now possible to determine the yrast energy. Since 
$L=N/2+m_{AB}$, the value of $m_{AB}$ is completely determined, 
and the minimum energy is obtained for the minimum value of 
$j_{AB}$.  For $0<L<N_B$, $-N/2<m_{AB}<-(N_A-N_B)/2$, and the 
minimum value of $j_{AB}$ is $|m_{AB}|=N/2-L$. Substituting this 
value of $j_{AB}$ into Eq.\,(\ref{eq:enereigen}) yields 
\begin{equation}
E_{0<L<N_B}^{\rm gr}=\frac{1}{2}\left( \frac{2\pi}{L_0} 
\right)^2L+\frac{\gamma_0}{L_0}\left( \frac{1}{2}N(N-1)-L\right). 
\label{eq:groundener1}
\end{equation}
The excited energy levels are given by Eq.\,(\ref{eq:enereigen}), 
with $N/2-L<j_{AB}\le N/2$.

For $N_B<L<N_A$ the minimum value of $j_{AB}$ is by $(N_A-N_B)/2$, 
independent of $L$. In this case the minimum energy is 
\begin{eqnarray}
E_{N_B<L<N_A}^{\rm gr}=\frac{1}{2}\left( \frac{2\pi}{L_0} \right)^2L+
\frac{\gamma_0}{L_0}\left(\frac{1}{2}N(N-1) \right.
\nonumber \\
\left.   -N_AN_B-N_B-L^2+NL\right). 
\label{eq:groundener2}
\end{eqnarray}
The energy levels of the excited states are given by 
Eq.\,(\ref{eq:enereigen}) with $(N_A-N_B)/2<j_{AB}\le N/2$.  

For $N_A<L<N$, $(N_A-N_B)/2<m_{AB}<N/2$, and the 
minimum value of $j_{AB}$ is $m_{AB}=L-N/2$. Substitution 
into Eq.\,(\ref{eq:enereigen}) now gives
\begin{equation}
   E_{N_A<L<N}^{\rm gr}=\frac{1}{2}\left( \frac{2\pi}{L_0} 
 \right)^2L+\frac{\gamma_0}{L_0}\left( \frac{1}{2}N(N-1)-N+L\right). 
\label{eq:groundener3}
\end{equation} 
The excited energy levels are again given by 
Eq.\,(\ref{eq:enereigen}) with $L-N/2<j_{AB}\le N/2$. 

\section{Conclusions}

The issue of finding solitary-wave solutions of the nonlinear
Schr\"odinger equation is an old problem with varying degrees 
of difficulty. The most elementary question is that of a single
component which extends to infinity. The case of a single component 
with periodic boundary conditions introduces some interesting 
complications. In the presence of a second component, the similar 
questions introduce additional complications, as there are now two 
coupled equations. Here we have considered the case of solitary-wave 
solutions in a two-component Bose-Einstein condensed gas, which 
is confined to a zero width (i.e., one dimensional) ring of finite 
radius, therefore requiring the imposition of periodic boundary 
conditions. 

Within the mean-field approximation and with the use of a reasonable 
ansatz for the solution, we have integrated the coupled nonlinear 
equations describing order parameters to find two analytic solutions 
which can be expressed in terms of Jacobi elliptic functions. 

This problem is also connected to the determination of the yrast
state, i.e., the state of lowest energy state solution given some
fixed value of the expectation value of the angular momentum.
We have shown explicitly that the yrast state for this problem 
is the gray-bright solution, in accordance with general arguments 
\cite{equiv}. The corresponding phase winding numbers that describe 
the global minima depend on the angular momentum of the system, 
giving rise in this way to various sectors of the dispersion 
relation, which nevertheless remains continuous. For weak 
coupling, this is shown analytically, however the numerical 
solution of the constraints suggests that the situation does 
not change qualitatively for stronger couplings. 

Going beyond the mean-field approximation, we have also 
diagonalized the many-body Hamiltonian exactly in the limit of 
weak interactions, which allows us to truncate the Hamiltonian 
to the two lowest-angular momentum modes. We have thus managed 
to derive the entire excitation spectrum of this many-body 
system, which in a sense corresponds to a ``quantum" 
solitary-wave solution.    

Ideally we would like to find solutions for arbitrary
masses $M_A$ and $M_B$, for arbitrary coupling constants
$\gamma_{AA}$, $\gamma_{BB}$, and $\gamma_{AB}$, and for 
arbitrary $u$. We have found analytic solutions by imposing 
the “artificial” constraints that $M_A = M_B$ and $\gamma_{AA} = 
\gamma_{BB} = \gamma_{AB}$. We expect that small violations of 
these constraints would lead to new (linear) equations that 
would be non-singular and well-behaved. This suggests that 
the present constrained solutions are broadly representative 
of all solutions which do not violate the constraints “violently”.  

\acknowledgements

This project is implemented through the Operational
Program ``Education and Lifelong Learning", Action
Archimedes III and is co-financed by the European Union
(European Social Fund) and Greek national funds (National
Strategic Reference Framework 2007 - 2013). We acknowledge 
support from the POLATOM Research Networking Programme of 
the European Science Foundation (ESF).

\section*{Appendix 1}

Let us consider the expansion of the gray-bright solution when 
$m$ is close to zero. The Lambert series  for the Jacobi elliptic 
functions tell us that, for $q(m)$ close to 1,
\begin{eqnarray}
{\rm sn}(\frac{2K(m)z}{L_0}|m)\sim \frac{2\pi }
{K(m)\sqrt{m}}\left(\frac{q(m)^{1/2}}{1-q(m)}\sin(\frac{\pi z}{L_0})
\right. \nonumber \\ \left.
+ \frac{q(m)^{3/2}}{1-q(m)^3}\sin(\frac{3\pi z}{L_0})+\cdots \right),
\label{apeq:lambert}
\end{eqnarray}
where $q(m)$ is the nome function. Its expansion for small $m$ is 
\begin{equation}
q(m)=\frac{m}{16}+\frac{m^2}{32}+{\cal O}(m^3).
\label{apeq:nome}
\end{equation}
Making use of the the expansion of the first elliptic integral $K(m)$, 
\begin{equation}
K(m)=\frac{\pi}{2}\left( 1+\frac{1}{4}m+\frac{9}{64}m^2+O(m^3)\right),
\label{apeq:ellipticfirst}
\end{equation}
we find that 
\begin{eqnarray}
{\rm sn}(\frac{2K(m)z}{L_0}|m)\sim \sin(\frac{\pi z}{L_0})+
\nonumber \\
+\frac{m}{16}
(\sin(\frac{\pi z}{L_0})+\sin(\frac{3\pi z}{L_0}))+{\cal O}(m^2).
\label{apeq:expsn}
\end{eqnarray}
Similarly, we find 
\begin{eqnarray}
{\rm cn}(\frac{2K(m)z}{L_0}|m)\sim \cos(\frac{\pi z}{L_0})+
\nonumber \\
+\frac{m}{16}(-\cos(\frac{\pi z}{L_0})+\cos(\frac{3\pi z}{L_0}))
+{\cal O}(m^2).
\label{apeq:expcn}
\end{eqnarray}
Expansions similar to Eq.\,(\ref{apeq:ellipticfirst}) also exist 
for the second and the third elliptic integrals,
\begin{equation}
E(m)=\frac{\pi}{2}\left( 1-\frac{1}{4}m
-\frac{3}{64}m^2+{\cal O}(m^3)\right),
\label{apeq:ellipticsecond}
\end{equation}
and 
\begin{eqnarray}
\Pi(a|m)=\frac{\pi}{2\sqrt{1-a}}+\frac{\pi m}{4a}\left(\frac{1}
{\sqrt{1-a}}-1\right)
\nonumber \\
-\frac{\pi m^2}{32a}\left(3a-\frac{6}
{\sqrt{1-a}}+6\right)+{\cal O}(m^3).
\label{apeq:ellipticthird}
\end{eqnarray}

\section*{Appendix 2}
We wish to evaluate the integral appearing in Eq.\,(\ref{eq:energy4}). 
In doing this we will use the integrals 
\begin{eqnarray}
\int_{-K(m)}^{K(m)}{\rm sn}^2(u|m)du&=&\frac{2(K(m)-E(m))}{m}, 
\\
\int_{-K(m)}^{K(m)}{\rm sn}^4(u|m)du&=&
\frac{2[(m+2)K(m)-2(m+1)E(m)]}{3m^2}.
\nonumber \\
\label{apeq:integ}
\end{eqnarray}
Recalling that the solution ansatz tells us that $n_B = \kappa 
n_A+\lambda$, the normalization condition gives $\lambda = (N_B
-\kappa N_A)/L_0$. Making this substitution in Eq.\,(\ref{eq:energy4}), 
we obtain 
\begin{eqnarray}
E_0=\gamma_0\Big{[}(1+\kappa )^2 \int_{-L_0/2}^{L_0/2} 
n_A^2 dz+\frac{N_AN_B}{L_0}-\frac{3}{2}\kappa \frac{N_A^2}
{L_0}
\nonumber \\
-\kappa^2\frac{N_A^2}{L_0}+\frac{N_B^2}{L_0}+\frac{1}
{2\kappa}\frac{N_B^2}{L_0}\Big{]}. \hspace*{2.0em}
\label{apeq:energy1}
\end{eqnarray}
The particle normalization condition can be used to reduce the 
remaining integral to 
\begin{eqnarray}
\int_{-L_0/2}^{L_0/2}{\rm sn}^2(\frac{2K(m)z}{L_0}|m)dz=\frac{L_0}{m}
(1-\frac{E(m)}{K(m)})
\nonumber \\
= \frac{N_A-n_{A ,{\rm min}}L_0}
{n_{A ,{\rm max}}-n_{A ,{\rm min}}}.\hspace*{8.0em}
\label{apeq:integ2}
\end{eqnarray}
The normalization condition also enables to eliminate the ratio 
$E(m)/K(m)$ to obtain 
\begin{eqnarray}
\int_{-L_0/2}^{L_0/2}{\rm sn}^4(\frac{2K(m)z}{L_0}|m)dz
\nonumber \\
= \frac{L_0}
{3m^2}\Big{[} (m+2)-2(m+1)\frac{E(m)}{K(m)}\Big{]}=
\nonumber \\
\frac{L_0}{3m}
\Big{[} -1+2(m+1)\frac{N_A/L_0-n_{A ,{\rm min}}}{n_{A ,{\rm max}}-
n_{A ,{\rm min}}}\Big{]}.
\label{apeq:integ3}
\end{eqnarray}
Substituting Eqs.\,(\ref{apeq:integ2}) and (\ref{apeq:integ3})
into Eq.\,(\ref{apeq:energy1}), we obtain Eq.\,(\ref{eq:energy5}).

\end{document}